\documentclass[12pt,a4paper]{article}
\usepackage{jheppub}
\usepackage[utf8]{inputenc}
\usepackage[english]{babel}
\usepackage{epsfig} 
\usepackage{amscd}
\usepackage{verbatim,amssymb,mathrsfs}
\usepackage{latexsym}
\usepackage{amsfonts,amsthm,amsmath,mathtools,amsmath,graphics}
\usepackage{appendix}

\notoc 

\title{$\tau_{RR}$ minimization in presence of hypermultiplets}
\preprint{}
\author[a]{Antonio Amariti}
\author[b]{\!\!,~Alessandra Gnecchi}
\affiliation[a]{INFN, Sezione di Milano, Via Celoria 16, I-20133 Milano, Italy}
\affiliation[b]{Max-Planck-Institut für Physik (Werner-Heisenberg-Institut), Föhringer Ring 6, 80805, München, Germany}

\emailAdd{antonio.amariti@mi.infn.it, agnecchi@mpp.mpg.de}

\abstract{We compute $\tau_{RR}$ minimization in gauged supergravity for M-theory and String Theory truncations with both massless and massive vector multiplets. We explicitly compute, as anticipated in \cite{Amariti:2015ybz}, that massive vector fields at the vacuum require the introduction of a constraint through a Lagrange multiplier. We illustrate this explicitly in two examples,  namely the $U(1)^2$-invariant truncation dual to the mABJM model and the ISO(7) truncation in massive IIA, the latter being a theory with both electric and magnetic gauging. We revisit the vacuum constraints at $AdS_4$ and show how the supergravity analysis matches the results of the field theory dual computation. }

\begin{document}

\begin{flushright}
    {MPP-2021-104}\\
    \end{flushright}

\maketitle

\newpage
\tableofcontents

\newpage
\section{Introduction}

The existence of a monotonic function that captures the irreversibility of the RG flow is an important issue in the study of quantum field theories. For conformal field theories, two theorems have been proved that ensure the existence of such a function in 2D \cite{Zamolodchikov:1986gt} and 4D \cite{Komargodski:2011vj}. 
The two functions that decrease at the endpoints of the RG flow are the
coefficient $c_{2D}$ of the Weyl anomaly for 2D CFTs and the coefficient of the Euler density $a_{4D}$ for  4D CFTs.
These  central charges that can be computed non-perturbatively in supersymmetric field theories, in terms of the exact $R$-charges. For minimal holomorphic supersymmetric field theories the $R$-current mixes with the abelian flavor symmetries and the exact knowledge of this mixing is necessary in order to compute the central charges.
The  mechanisms to determine the coefficients of this mixing have been obtained in \cite{Benini:2012cz,Benini:2013cda} in the 2D case and in  \cite{Intriligator:2003jj}  for the 4D case. In the first case the central charge $c_{2D}$  is extremized 
by the exact $R$-charge at the superconformal point while in the second case the central charge $a_{4D}$ is maximized by the exact $R$-charges at the superconformal fixed point.

In odd dimensions one cannot rely on anomalies. However, in 3D SCFTs, looking at the coefficient of the two point function of the stress tensor it was shown that it is possible to use $\tau_{RR}$ to determine the exact R-current \cite{Barnes:2005bm}. The non-perturbative contributions to this two point functions, however, are not known, so most of the work on extremization has been focused instead on the 3D SCFT free energy which, thanks to localization techniques, can be computed non-perturbatively. In fact, for 3D SCFT it has been shown that the realized R-symmetry is the charge assignment that maximizes the free energy of the theory on the sphere $S^3$ \cite{Jafferis:2010un}, making it a reliable quantity to determine the R-symmetry at the fixed point. 

In the context of holography, it is interesting to understand what are the gravitational counterparts of the various mechanisms that determine the R-symmetry in the dual field theory, at the superconformal fixed point. It is well established now that, in various dimensions, the attractor mechanism on the supergravity scalars (that determine their value at the AdS vacuum) can be used holographically to determine the R-symmetry realized in the SCFT.  In the context of $AdS_5/CFT_4$, this has been discussed in \cite{Tachikawa:2005tq} while for $AdS_4/CFT_3$ the discussion has been generalized in \cite{Amariti:2015ybz}.

With this note, we want to explicitly illustrate the mechanism of extremization of the $\tau_{RR}$ function holographically for $AdS_4$ vacua in N=2 supergravity theories coupled with hypermultiplets, looking explicitly at  two models of 3D SCFT with a known holographic dual description and with a consistent truncation that has enough vector multiplets to reproduce the mixing of the $R$-charge  with the abelian global currents.

The presence of hypermultiplets in the gravity theories allows to impose the superpotential constraints of the 3D theory, thus playing the role of Lagrange multipliers, as proposed in \cite{Amariti:2015ybz}. Observe that recent works on black holes in $AdS_4$ have explicitly computed these constraints \cite{Benini:2017oxt,Hosseini:2017fjo}. Analogously in the $AdS_5/CFT_4$ context, this mechanism had been proposed in \cite{Tachikawa:2005tq} and explicitly analysed in \cite{Szepietowski:2012tb}. 

The paper is organized as follows. We review the principle of $\tau_{RR}$ minimization and its field theory realization in Section \ref{sec-2}, concentrating on the models that will be relevant for the rest of the paper. In Section \ref{sec-sugra} we review the properties of $AdS_4$ vacua of 4d gauged N=2 supergravity and we prescribe their holographic interpretation in relation to $\tau_{RR}$-minimization, extending to the case of hypermultiplet the study of \cite{Amariti:2015ybz}. In Section \ref{sec-U1squared-trunc} we explicitly work out the case of the N=2 gauged supergravity model obtained as an $U(1)^2$-invariant truncation of N=8 supergravity, which is dual to massive deformation of ABJM (mABJM). We study the minimization in presence of hypermultiplets as a constraint extremization mechanism thanks to appropriate Lagrange multipliers, and we show how the R-charges are identified holographically, matching the field theory result. In Section \ref{sec-ISO7-trunc} we focus instead on a massive IIA vacuum and show that even in the case of dyonic gauging our prescription allows to match the supergravity analysis with the field theory computation. We conclude with an outlook on possible extensions of these results to other theories and known obstructions thereof.

\section{$\tau_{RR}$ extremization and its field theory derivation \label{sec-2}}

In this section we review some basic aspects about  the behavior  $\tau_{RR}$.
The function $\tau_{RR}$  parameterizes the mixing of the UV $R$-current 
with the global symmetries.
 The $\tau_{RR}$ function expressed in terms of the mixing of the UV $R$-current 
with the global symmetries  is referred as 
\emph{off-shell} and the exact IR combination is determined by minimizing 
this \emph{off-shell}  function.
Such a  minimization principle holds for generic dimensionality
and this property parallels the volume minimization 
 of the gravitational dual picture. Indeed it has been shown that 
$\tau_{RR}$ is proportional at large $N$ to the volume,
and that $\tau_{RR}$-minimization is equivalent to volume minimization \cite{Barnes:2005bw}.

A full quantum computation of $\tau_{RR}$ for a 3d SCFT requires to take into account the 
effects from the loop expansion and the calculation is rather complicated. Furthermore 
for strongly coupled field theories a non-perturbative analysis is necessary.
Anyway, as shown in \cite{Closset:2012ru}, $\tau_{RR}$ corresponds to the three sphere free energy 
at large $N$.
The precise relation between the free energy and the $\tau_{RR}$ function at the fixed point is
\begin{equation}
\tau_{RR}^{min} = \frac{\pi^2}{4} F_{S^3}^{max} 
\end{equation}
where  $\tau_{RR}^{min}$ and $F_{S^3}^{max}$ are the $\tau_{RR}$ function and  the free energy at the superconformal fixed point respectively.
Furthermore for models with a gravitational dual description one can use the relations between the 
$\tau_{RR}$ function and the volume to infer the off-shell relation between $\tau_{RR}$ and the free energy.
It reads
\begin{equation}
\tau_{RR}(\Delta) = \frac{\pi^2}{4} \frac{(F_{S^3}^{max})^3}{F_{S^3}^2(\Delta)} 
\end{equation}
where $\tau_{RR}(\Delta) $ and $F_{S^3}(\Delta)$ correspond to the coefficients of the two point function for the R-current correlator and to the free energy in terms of the mixing of the R-current with the global symmetries respectively.
Here we concentrate on the calculation of $\tau_{RR}$ for models with a conjectured gravity dual description, and this allows us to use the  calculation of the free energy on $S^3$ at large $N$.
Here we focus on two models with a conjectured gravity dual in order to match the field theory computations with the ones from supergravity.
The first models that we discuss is the so called mABJM theory, a massive deformation of ABJM through a monopole superpotential.
The second theory is  a CS gauge theory conjectured in \cite{Guarino:2015jca} to be dual
massive type IIA supergravity on a manifold with the topology of $S^6$.
Let's quickly review these models and their $\tau_{RR}$ functions in terms of the mixing of the abelian global currents with $U(1)_R$.
\subsection{The mABJM theory}

We start our analysis with the mABJM theory studied in 
\cite{Benna:2008zy,Klebanov:2008vq}.
This models corresponds a deformation of the ABJM theory, with a 
deformed superpotential given by
\begin{equation}
\label{wdef}
W = \epsilon_{ij} \epsilon_{lk} A_i B_l A_j B_k +
T^{(1)} A_1
\end{equation}
where the traces are understood.
The fields  $A_i$ and $B_i$ are respectively the bifundamentals 
and the antibifundamentals of the $U(N) \times U(N)$ 
gauge group.
The monopole operator in the deformation is referred as  $T^{(1)}$,
with unit of flux through an $S^2$
surrounding the insertion point
turned on for the topological symmetry.
The second term in (\ref{wdef})  
drives an RG flow through an interacting $\mathcal{N}=2$ fixed point.
At large $N$ the three sphere  free energy for this IR SCFT has been computed in  \cite{Jafferis:2011zi}
and it is
\begin{equation}
F_{S^3} = \frac{4 \sqrt 2 \pi}{3}
 N^{3/2} \sqrt{\Delta_{A_1} \Delta_{A_2}\Delta_{B_1} \Delta_{B_2} }
\end{equation}
with the constraints imposed by the superpotential corresponding to 
$\Delta_{A_1} = 1$ and $\Delta_{A_2}+\Delta_{B_1} + \Delta_{B_2} = 1$.
We can parameterize the charges as
\begin{eqnarray}\label{mABJM-charges}
&&
\Delta_{A_1} =  \frac{1}{2} (\delta_0 + \delta_1+\delta_2 +\delta_3),
\quad
\Delta_{A_2} =   \frac{1}{2} (\delta_0 + \delta_1-\delta_2 -\delta_3)
\nonumber \\
&&
\Delta_{B_1} =   \frac{1}{2} (\delta_0 - \delta_1+\delta_2 -\delta_3)
, \quad
\Delta_{B_2} =   \frac{1}{2} (\delta_0 - \delta_1-\delta_2 +\delta_3)
\end{eqnarray}
where the exact $R$-symmetry is given by 
$U(1)_R = \sum_i \delta_i U(1)_i$.
Imposing the constraints on $\Delta_i$ we have $\delta_0=1$ and
$\delta_1+\delta_2+\delta_3=1$.
The free energy and the function $\tau_{RR}$
 in terms of the mixing parameters are
\begin{equation}
\label{ftres}
F = \frac{4 \sqrt 2 \pi}{3}
 N^{3/2} \sqrt{\delta_1 \delta_2  \delta_3 },
 \quad
 \tau_{RR} =\frac{ N^{3/2}}{3 \pi }  \left(\frac{2}{3}\right)^\frac{9}{2}
\frac{1}{\delta_1 \delta_2 \delta_3 }
 \ .
 \end{equation}

\subsection{The SCFT dual to massive IIA on $S^6$}

This second model corresponds to an $U(N_c)$ gauge theory at CS level $k$ with three adjoints 
$X,Y$ and $Z$ interacting through the superpotential 
\begin{equation}
W = X[Y,Z]
\end{equation} 
This theory has been conjectured in \cite{Guarino:2015jca} to be dual to massive IIA compactified on a manifold with the topology of $S^6$.
The model has $\mathcal{N}=3$ supersymmetry and $SU(3) \times U(1)_R$ global symmetry.
The $U(1)_R$ $R$-charges of the fields are fixed at the superconformal value $\frac{2}{3}$.
Anyway we can parameterize the charges of the fields as $\Delta_X$, $\Delta_Y$ and $\Delta_Z$ 
constrained by $\Delta_X + \Delta_Y + \Delta_Z=2$
and then compute the free energy at large $N$ as done in \cite{Fluder:2015eoa}.
The free energy and the function $\tau_{RR}$
 in terms of these charges parameters are
\begin{equation}
\label{fluder}
F_{S^3} = \frac{9  3^{\frac{1}{6}} \pi  \left(\Delta _X \Delta _Y \Delta _Z\right)^{2/3} k^{\frac{1}{3}} N^{\frac{5}{3}}}{10 2^{\frac{2}{3}}}\ ,
 \quad
\tau_{RR} = \frac{64 \sqrt[3]{2} \sqrt[3]{k} N^{5/3}}{135\ 3^{5/6} \pi  \left(\Delta _X \Delta _Y \Delta _Z\right){}^{4/3}} \ .
\end{equation}
\\
In the following sections we will reproduce (\ref{ftres}) and (\ref{fluder}) from the consistent truncations of the conjectured gravitational dual descriptions.

\section{N=2 Supergravity and $AdS_4$ vacua revisited \label{sec-sugra}}

We are interested in $N=2$ vacua with nonzero cosmological constant, solutions of 4d \emph{gauged} supergravity\footnote{We follow the formulation of N=2 4d gauged supergravity of \cite{Andrianopoli:1996cm}. Vacuum conditions for supersymmetric AdS vacua of 4d gauged supergravity have been studied extensively in \cite{Hristov:2009uj,Louis:2012ux,Erbin:2014hsa}.}. We will focus on known examples whose origin in M-theory and String Theory is known. 

\subsection{Vacua of theories with only vector multiplets}

The simplest example one can consider corresponds to a vacuum of $N=8$ supergravity, obtained as an M-theory reduction on $AdS_4\times S^7$, dual to 3d ABJM theory \cite{Aharony:2008ug}. When we further truncate the N=8 theory to $N=2$, the $U(1)\subset SU(2)_R$ of R-symmetry is gauged. This is the simplest gauging, since it does not require the presence of hypermultiplets. This vacuum can be obtained in fact as a maximally supersymmetric solution of $N=2$ supergravity coupled to 4 vector multiplets with Fayet-Iliopouls gauging \cite{Cacciatori:2009iz,Freedman:2013ryh}. 

The effective theory around this vacuum is an $N=2$ supersymmetric theory coupled to three abelian, massless vector multiplets described by the bosonic lagrangian 
    \begin{align}\label{SGaction-intext}
    S&=\int d^4 x\left(- \frac R2+g_{i\bar{j}}\partial_{\mu}z^i\partial^{\mu}\bar z^{\bar j} + V(z^i, \bar z^{\bar i}, \mathcal{G})
    \right.
    \nonumber
    \\
    &\hspace{2cm}
    \left.
    +
    \mathcal{I}_{\Lambda \Sigma}F^{\Lambda}_{\mu \nu}F^{\Sigma\,\mu\nu} +\frac1{2\sqrt{-g}}\mathcal{R}_{\Lambda \Sigma}\epsilon^{\mu \nu \rho \sigma}F^{\Lambda}_{\mu \nu}F^{\Sigma}_{\rho \sigma}
    \right)\ .
    \end{align}
The gauging parameters $\mathcal{G}$ determine the superpotential\footnote{The explicit form of this expressions from supergravity is left to the appendix.
}
    \begin{align}
        W(z,\bar z, \mathcal{G})  &\equiv  \mathcal{L} =\langle \mathcal{G}\,, \mathcal{V}\rangle \ ,
    \end{align}
and thus the scalar potential 
    \begin{eqnarray}
    V_g(z,\bar z, \mathcal{G})&=&
    -3 W\overline{W}+g^{i\bar j}D_i W D_{\bar j}\overline{W}\  .
    \end{eqnarray}
The scalar potential yields the cosmological constant at the vacuum, and can only be nonzero in presence of gauging, thus $\mathcal{G}\neq 0$. Supersymmetry requires that at the vacuum the gauge fields are zero and the scalars are fixed by the condition 
    \begin{align}\label{attr-ABJM}
        D_i \mathcal{L} =0\ .
    \end{align}

A generic $AdS_4$ vacuum, however, will be obtained by more general gaugings of the isometries of the scalar manifolds. In principle, both vector multiplets scalars as well as hypermultiplets are coordinates of scalar manifolds whose isometries can be gauged. When the gauged isometries correspond to an abelian group, only a subgroup of isometries of the hypermultiplet scalar manifolds are gauged. 

The $\tau_{RR}$ minimization for the vacuum discussed so far was performed in \cite{Amariti:2015ybz}, where it was shown that the R-symmetry realization via mixing among the dual flavor symmetries is determined by the attractor condition \eqref{attr-ABJM}.

In addition to this, for generic vacua, we will show  that non-vanishing hypermultiplets serves as Stuckelberg fields and in fact, in the examples of this work, the $AdS_4$ vacua will be solution of an effective $N=2$ supergravity theory coupled to massless and \emph{massive} vector multiplets.

\subsection{Vacua of theories with hypermultiplets}

When the gauging involves dynamical hypermultiplets, the bosonic action reads 
\begin{align}\label{SGaction-intext-hypers}
    S&=\int d^4 x\left(- \frac R2+g_{i\bar{j}}\nabla_{\mu}z^i\nabla^{\mu}\bar z^{\bar j}+h_{uv}\nabla_\mu q^u\nabla^\mu q^v-V_g(z,\bar z,q)
    \right.
    \nonumber
    \\
    &\hspace{2cm}
    \left.
    +
    \mathcal{I}_{\Lambda \Sigma}F^{\Lambda}_{\mu \nu}F^{\Sigma\,\mu\nu} +\frac1{2\sqrt{-g}}\mathcal{R}_{\Lambda \Sigma}\epsilon^{\mu \nu \rho \sigma}F^{\Lambda}_{\mu \nu}F^{\Sigma}_{\rho \sigma}
    \right)\ ,
    \end{align}    
and the scalar potential takes the general form
\begin{eqnarray}\label{scalar-V-gauged-SG}
V_g(z,\bar z, q)&=&4 h_{uv}\langle k^u(q),\,\mathcal{V}(z,\bar z) \rangle\, \langle k^u(q),\,\overline{\mathcal{V}}(z,\bar z)\rangle
-3 \mathcal{L}^x\overline{\mathcal{L}^x}+g^{i\bar j}D_i \mathcal{L}^x D_{\bar j}\overline{\mathcal{L}^x}\ , \ \ 
\end{eqnarray}
(with sum over $x$ indices). This is expressed in terms of a triplet of functions $\mathcal{L}^x$ \footnote{We will generically consider both electric and magnetic gauging, thus we indicate
\begin{eqnarray} \label{embedding-tensor-to-Pcal-Kill}
\mathcal{P}^x=\mathcal{P}^x_\lambda(\Theta^{\Lambda \lambda},\Theta^\lambda_\Lambda)\ ,
\qquad
k^u(q)=k^u_\lambda(\Theta^{\Lambda \lambda},\Theta^\lambda_\Lambda)\ ,
\end{eqnarray}
for a generic choice of embedding tensor $\Theta$ \cite{deWit:2011gk}.}
\begin{align}\label{W-def}  
\mathcal{L}^x(z\,, \bar z\,, q^u) = \langle \mathcal{P}^x(q),\, \mathcal{V}(z,\bar z)\rangle\equiv e^{K/2}\left(\mathcal P^x_{\Lambda}X^{\Lambda}-\mathcal P^{x\Lambda} F_{\Lambda}\right) \ .
\end{align}

Notice that, if no isometry of the Special K\"ahler manifold is gauged (i.e. no scalar of the vector multiplets are charged under the vectors), then the hypermultiplets scalars can be charged only under abelian gauge fields.
For gauging involving isometries of the quaternionic manifold only, then, Supersymmetry requires that the gauge fields are zero at the vacuum \cite{Hristov:2009uj} and that the scalars obey the following equations 
    \begin{align}
        \label{AdS4-hyp}  
        D_i \mathcal{L}^x(z\,, \bar z\,, q^u) &= 0 \ ,
        \qquad
            \langle k^u(q^*),\,\mathcal{V}(z^*,\bar z^*) \rangle = 0\ .
    \end{align}
Supersymmetry also set the AdS radius to be 
\begin{align}\label{AdS4-radius}
    |\mathcal{L}^x(z^*,\bar z^*,q^{u\,*})|^2=\frac{1}{\ell^2_{AdS}} \ .
\end{align}
Clearly the $U(1)$ R-symmetry gauging corresponds to the case of constant tri-holomorphic moment maps $\mathcal{P}^x=(\mathcal{P}^{\Lambda\,x},\mathcal{P}_\Lambda^x)$. In this case, moreover, since the Killing vectors are identically zero, it is possible to solve the equivariant condition \eqref{eq:equivariant} for a triplet of moment maps where $\mathcal{P}^x=\delta^{x3}\mathcal{P}^3$ off shell, for example, so that one retrieves the potential \eqref{W-def} by simply taking $\mathcal{P}^3\equiv \mathcal{G}$. From the same equivariant relation \eqref{eq:equivariant}, it is clear that it's not possible to rotate the moment maps to a single non-vanishing component generally off-shell but, in case of abelian gaugings, it is possible to gauge away two of the three component of the $SU(2)$ vector of moment maps at the vacuum.

Whenever the quaternionic geometry is a symmetric space with compact isotropy group, the quaternionic metric is positive definite, and the second vacuum condition in \ref{AdS4-hyp} is equivalent to
    \begin{align}
        \langle k^u_*,\,\mathcal{V}_* \rangle
        \langle k^v_*,\,\overline{\mathcal{V}}_* \rangle h_{*uv}&=0 \ .
    \end{align}
It is important to derive, using properties of the quaternionic geometry\footnote{See appendix \ref{subsec:quatgeo} for the derivation.} the following identity
    \begin{align}
        h_{uv}k^u k^v & = - \frac{1}{3\lambda^2}\sum_x \nabla^u\mathcal{P}^x\nabla_u\mathcal{P}^x \ , 
    \end{align}
which allows us to re-write the vacuum condition as\footnote{Supersymmetry requires $\lambda^2=1$.}
    \begin{align}
        \langle \nabla_u\mathcal{P}^x,\mathcal{V}\rangle =\nabla_u\mathcal{L}^x = 0 \ , \qquad 
        \forall x\in\{1,2,3\}\ .
    \end{align}
We can use this identity also in the action \ref{SGaction-intext} to re-write the first term of the scalar potential \ref{scalar-V-gauged-SG}
    \begin{align}
        \label{scalar-V-momentmaps}
        4 h_{uv}\langle k^u(q),\,\mathcal{V}(z,\bar z) \rangle\, \langle k^u(q),\,\overline{\mathcal{V}}(z,\bar z)\rangle=&-\frac{4}{3}  h^{uv}\langle \nabla_u\mathcal{P}^x(q),\,\mathcal{V}(z,\bar z) \rangle\, \langle \nabla_v\mathcal{P}^x(q),\,\overline{\mathcal{V}}(z,\bar z)\rangle 
        \  ,
    \end{align}
which can be rewritten in terms of an $SU(2)$ triplet $\vec{\mathcal{L}}=(\mathcal{L}^x)_{x=1,2,3}$ as \cite{Klemm:2016wng}
    \begin{align}
        \label{scalar-V-Lcalx}
        V_g(z,\bar z, q)&=\sum_x\left(-\frac{4}{3}  h^{uv}\nabla_u\mathcal{L}^x\nabla_v\overline{\mathcal{L}}^x(q^u,z^i,\bar z^{\bar i})
-3 \mathcal{L}^x\overline{\mathcal{L}^x}+g^{i\bar j}D_i \mathcal{L}^x D_{\bar j}\overline{\mathcal{L}^x}
        \right)\  .
    \end{align}
Notice that the vacuum conditions have further implications, due to the equivariant relation. In fact, if we contract \eqref{eq:equivariant} with the symplectic sections, we obtain that on the vacuum 
    \begin{align}
        \frac12 \epsilon^{xyz}\mathcal{L_*}^z\overline{\mathcal{L_*}}^z+ f^\Gamma_{\Lambda\Sigma}P^x_\Gamma L_*^\Lambda\overline{L_*}^\Sigma = 0 \ ,
    \end{align}
but by consistency of the gauging one has \cite{DAuria:1990qxt}  $f^\Gamma_{\Lambda\Sigma} L^\Lambda\overline{L}^\Sigma = 0$ identically, so at the vacuum the sections $\mathcal{L}^x$ satisfy\footnote{This condition was found in \cite{Hristov:2009uj} as consequence of an integrability condition on the supersymmetry variation. We have shown here that this can be derived as a consequence of the equivatiant condition.}
    \begin{align}
        \epsilon^{xyz}\mathcal{L_*}^z\overline{\mathcal{L_*}}^z = 0.
    \end{align}
Using the special geometry identity \eqref{SG-id-1} this condition is equivalent to 
    \begin{align}\label{locality}
        \langle \mathcal{P}_*^x, \mathcal{P}_*^y\rangle =0\ ,
        \qquad 
        \forall \ x,y=1,2,3 \ .
    \end{align}
This constraints is just the locality condition on the electric and magnetic hypermultiplet charges, corresponding to a locality condition on the electric and magnetic components of the embedding tensor \cite{deWit:2007kvg}. 
We can summarize the vacuum condition for a fully supersymmetric $AdS_4$ solution of gauged N=2 supergravity as
    \begin{align}\label{attr-hyp}
        D_i \mathcal{L}^x(z\,, \bar z\,, q^u) &= 0 \ ,
        \qquad 
        \nabla_u \mathcal{L}^x (z\,, \bar z\,, q^u)=0 \ ,
        \nonumber \\
        \langle \mathcal{P}_*^x, \mathcal{P}_*^y\rangle =0\ ,
        & \qquad 
        \forall \ x,y=1,2,3 \ ,
        \nonumber
        \\
        \big|\sum_x\mathcal{L}_*^x\overline{\mathcal{L}}_*^x\big|&= \frac{1}{\ell^{2}_{AdS}} \ .
    \end{align}
It is known that, using special geometry identities, the first equations in \eqref{attr-hyp} can be solved as
    \begin{align}\label{Pvac}
        \mathcal{P}^x_* &= - c^x \frac{2}{\ell_{AdS}}\textrm{Im}(e^{-i\psi}\mathcal{V}) \ ,
    \end{align}
for a constant SU(2) vector $c^x$.
Let us consider one last time the second equation of \eqref{attr-hyp} 
    \begin{align}\label{extrem-hyp-Killing}
        \langle k^u(q^*),\,\mathcal{V}(z^*,\bar z^*) \rangle = 0\ .
    \end{align}
This equation is nontrivial for those $k^u_\lambda$ that are nonzero at the vacuum. Since the supergravity theory is an abelian gauge theory, the Killing vectors that are nonzero at the vacuum signal the spontaneous breaking of its corresponding isometry, with the consequence that the vector field gauging this isometry becomes massive . Since the isometries of the scalar manifolds are embedded in the symplectic $Sp(n_v+1,\mathbb{R})$ group via the embedding tensor $(\Theta^{\lambda\Lambda},\Theta^\lambda_\Lambda)$, the number of massive gauge fields is given by \cite{Louis:2012ux}
    \begin{align}
        n=\textrm{rk}(\Theta^\lambda k^u_\lambda)=
        \textrm{rk}\left(\begin{array}{c}  
            \Theta^{\lambda\Lambda}k^u_\lambda(q^*)
            \\
            \Theta^{\lambda}_\Lambda k^u_\lambda(q^*)
        \end{array}\right)
    \end{align}
As pointed out in \cite{Louis:2012ux} from a general analysis of the vacuum constraints, the massive vectors belong to long multiplets
    \begin{align}
        1(1) \ \ 2(1/2) \ \ 5(0) 
    \end{align}
obtained from one massless vector multiplet and one hypermultiplet of the unbroken theory. 
The identification of the Higgsing is important since when the $AdS_4$ vacuum is studied in a holographic context, what matches the dual field theory is the effective supergravity theory around the $AdS_4$ vacuum, where the vectors have been Higgsed. 
\subsection{Holographic matching}
The goal of this work is to relate the above analysis of $AdS_4$ vacua holographically to the field theory R-symmetry charges realized in the dual SCFT, thanks to the extremization of the $\tau_{RR}$ coefficient of the stress tensor two point function. 

In our work \cite{Amariti:2015ybz}, we have proven how the $\tau_{RR}$ function 
    \begin{align}\label{tau-RR}
        \tau_{RR}&=\frac{4}{\pi^2} F_{S^3}^{max}\frac{|W|^4}{\mathcal I_4} \ ,
    \end{align}
is minimized at the vacuum\footnote{Here, $F_{S^3}^{max}$ is the free energy of the dual field theory, $\mathcal{I}_4(\mathcal G)$ is a symplectic invatiant of the special K\"ahler scalar manifold, evaluated on the Fayet-Iliopoulos parameters $\mathcal{G}^T=(g^\Lambda,g_\Lambda)$, and $W(z^i,\bar z^{\bar i}, q^u)$ is the superpotential $W=\langle \mathcal{G}, \mathcal{V}\rangle$. At the $AdS_4$ values of the scalars the superpotential coincides with the quartic invariant: $\mathcal{I}_4=W\Big|_{(z^i_*, \bar z_*^{\bar i}, q_*^u)}.$} as a function of the trial R-charges. 

The main ingredient of the holographic matching is the identification of the R-charges $s^I$ in the supergravity picture. We have proposed this identification in the case of vacua with no hypermultiplets turned on (i.e. only massless vectors) in \cite{Amariti:2015ybz}. We will show that even in presence of hypermultiplets this prescription holds, namely
    \begin{align}
        s^I &= e^{\mathcal{K}/2} \frac{X^I(z^i)}{\mathcal{W}} \ ,
    \end{align}
but this has to be supplemented by a constraint on the scalars $z^i$ that has to be imposed together with this identification. Notice that R-charges are associated only to massless vector fields, so the physical meaning of the constraints is to eliminate the \emph{unphysical} $s^I$ corresponding to vectors becoming massive at the $AdS_4$ vacuum.

Let's now move to the explicit analysis of two models that will illustrate how the matching works for theories with electric as well as dyonic gauging.

\section{ $U(1)^2$-invariant truncation of N=8 gauged supergravity \label{sec-U1squared-trunc}}

The simplest example of a vacuum with nontrivial hypermultiplet scalar, allowing us to clarify the above explanation, is an $AdS_{4}$ vacuum preserving $SU(3)\times U(1)$ symmetry in N=8 supergravity \cite{Warner:1983vz}. It is of our interest since it can be constructed in the N=2 truncation of N=8 \cite{Bobev:2010ib}, which consists of $n_V=3$ vector multiplets, like the previous example, but with in addition one hypermultiplet $n_H=1$, whose scalar manifold is
    \begin{align}
        \mathcal M_{SK} \times \mathcal M_{\mathbb Q}&= \left[
            \frac{SU(1,1)}{U(1)}    
        \right]^3
        \times
        \frac{SU(2,1)}{U(2)} \ .
    \end{align}
This N=2 theory can be seen as a truncation of the N=8, $SO(8)$-gauged Supergravity \cite{deWit:1982bul} to its $U(1)^2$ invariant sector. The details of this relation are explained in the Appendix, here we will work within the N=2 formulation.

The Special K\"ahler manifold $ \left[
    \frac{SU(1,1)}{U(1)}    
\right]^3$ can be parametrized by the metric\footnote{See Appendix \ref{append-spec-geom} for  details on Special Geometry notation.}
\begin{align}
    ds^2&=g_{i\bar \jmath} dz^i d\bar z^{\bar\jmath} = \sum_{i=1}^3 \frac{dz^i d\bar z^{\bar\imath}}{(z^i+\bar z^{\bar\imath})^2} \ ,
\end{align}
which is a K\"ahler metric obtained from the holomorphic prepotential $F(X^\Lambda)=-2i\sqrt{X^0 X^1 X^2 X^3}$ in terms of the projective coordinates $X^\Lambda(z)$. We fix the symplectic gauge by choosing
    \begin{align}
        X^0=1\ , \qquad X^1= z^2z^3 \ , \qquad X^2 = z^1 z^3 \ , 
        \qquad  X^3 = z^1 z^2 \ ,
    \end{align}
and in these coordinates the K\"ahler potential is 
    \begin{align}
        K= - \log\left[
        (z_1+\bar z_1)(z_2+\bar z_2)(z_3+\bar z_3)    
        \right] \ .
    \end{align}
The interesting aspect is that, in comparison to the setup of \cite{Amariti:2015ybz}, in this theory there is an additional quaternionic scalar manifold corresponding to four additional real scalars, which parametrize $SU(2,1)/U(2)$. This non linear sigma model is called Universal Hypermultiplet and is known in the literature in several parametrizations. The truncation from N=8 naturally yields complex coordinates $\zeta_1,\zeta_2$ and the K\"ahler metric of $SU(2,1)/U(2)$, as described in appendix \ref{app-truncation}. The transformation 
    \begin{align}
        \frac{\zeta_2}{1+\zeta_1}&= \theta - i\tau \ ,
        \qquad 
        \frac{1-\zeta_1}{1+\zeta_1} = V+(\theta^2+\tau^2)+i \sigma \ ,
    \end{align}
relates the K\"ahler coordinates to the quaternionic $q^u=(V,\sigma,\theta,\tau)$ \cite{Ceresole:2001wi} in terms of which the metric becomes
    \begin{align}\label{metric-Univ-Hyp}
        ds^2&=\frac{dV^2}{2V^2}+\frac{1}{2V^2}(d\sigma+2\theta d\tau-2\theta d\tau)^2+\frac2V(d\tau^2+d\theta^2) \ .
    \end{align}
The gauging of the N=2 theory correspond to a $U(1)^2$ subgroup of isometries of the N=8 theory. Their action can easily be expressed in terms of the K\"ahler coordinates as the rotations of the phases of $\zeta_1\,,\zeta_2$, with Killing vectors
    \begin{align}
        k_{(i)}& = i(\zeta_1\partial_{\zeta_1}+\zeta_2\partial_{\zeta_2}) +c.c.
        \nonumber \\
        k_{(ii)} & = i(\zeta_1\partial_{\zeta_1}-\zeta_2\partial_{\zeta_2})+c.c.
    \end{align}
In the N=2 formulation the Killing vectors are recast in a symplectic vector
\begin{align}\label{Killing-body}
    k^u=(k^{\Lambda\,u},k_\Lambda^u) \ ,  \qquad k_\Lambda^u=(k_0^u,k_1^u,k_1^u,k_1^u) \ , \qquad k^{\Lambda u}=0 \ \ ;
\end{align}
and their corresponding moment maps
\begin{align}
    \mathcal{P}^x=(\mathcal{P}^{x,\Lambda},\mathcal{P}^x_\Lambda)\ , 
    \qquad \mathcal{P}^x_\Lambda = (\mathcal{P}^x_0,\mathcal{P}^x_1,\mathcal{P}^x_1,\mathcal{P}^x_1) \ ,
    \qquad \mathcal{P}^{x\,\Lambda}=0 \ .
\end{align}
The explicit form of $(k^u_0\,,\,k^u_1)$ and $(\mathcal{P}^x_0,\mathcal{P}^x_1)$ is presented in \eqref{Kill-V-Coord} and \eqref{Pcal-V-coord} of appendix \ref{app-truncation}.

Notice that, in going from the Killing vectors of the $U(1)^2$ isometries in the coordinates of \eqref{metric-Univ-Hyp}, to the definition of the symplectic Killing vector $k^u=(k^{\Lambda\,u},k_\Lambda^u) $ in eq. \eqref{Killing-body}, we have to chose a symplectic embedding of the isometries that is actually compatible with the formulation of the supergravity theory. 
This choice is defined by the already mentioned embedding tensor $\Theta^\lambda=(\Theta^{\lambda\,\Lambda},\Theta^\lambda_\Lambda)$ as in \eqref{embedding-tensor-to-Pcal-Kill}, provided this tensor satisfies the constraints of \cite{deWit:2011gk}. 

Analogously, for a purely electric gauging like the one of this model, a valid embedding tensor is such that the equivariant condition \eqref{eq:equivariant} is satisfied by the chosen $k_\Lambda$ and $\mathcal{P}^x_\Lambda$. 

Actually, since we are dealing with an abelian isometry group, the equivariant condition is satisfied for any choice of embedding\footnote{We exclude the choice of $\Theta^\lambda$ for which $\text{rk}(\Theta^\lambda)<n_V+1$ also outside of the vacuum, since that actually gauges one only of the two isometries}. Different choices in fact simply correspond to a different parametrization of the special K\" ahler manifold and thus can be related by a special geometry fields redefinition. Different embedding tensors will simply differ by the vacuum values of the scalar $z^i$. Our choice can be read from  \eqref{Kill-V-Coord} and \eqref{Pcal-V-coord} and corresponds to
\footnote{This choice corresponds to the embedding tensor
    \begin{align}
        \Theta^\lambda&=\left(\begin{array}{cc}
            \Theta^{(i)\Lambda} & \Theta^{(ii)\Lambda} 
            \\
            \Theta^{(i)}_{\Lambda} & \Theta^{(ii)}_{\Lambda} 
        \end{array}\right) =
        \left(\begin{array}{cc}
            0 & 0 
            \\
            \vec{0} & \vec{0}
            \\
            -1 & 0
            \\
            \vec{2} & \vec{1}
        \end{array}\right)
    \end{align}
however, this is not the tensor that enters in the extremization mechanism, as it will become clear in what follows.
}:
    \begin{align}
        k^u_0& = - k^u_{(i)}
        \qquad 
        k^u_1 = 2 k^u_{(i)}+k^u_{(ii)} \ ,
        \qquad 
        k^u_1=k^u_2=k^u_3 \ .
    \end{align}

Now that we have fully determined the theory, the vacuum conditions \eqref{AdS4-hyp} or \eqref{attr-hyp} fix the scalars to the $AdS_4$ values\footnote{Notice there is actually a one-parameter family of vacua for $\langle\theta^2+\tau^2\rangle=\frac13$ but we focus on one realization for our analysis.}
    \begin{align}
        \langle z_i \rangle=\frac{1}{\sqrt{3}}\ , \quad \forall \ i=1,2,3\ ,
        \qquad
        \langle V \rangle = \frac{2}{\sqrt3} \ ,
        \qquad 
        \langle \theta \rangle = \frac{1}{\sqrt{3}} \ ,
        \qquad \langle \tau \rangle = \langle \sigma \rangle = 0  \ .
    \end{align}
It is easy to see, now, that at the vacuum one Killing vector field is zero
    \begin{align}
        k_{nonH}=k_{(i)}+k_{(ii)} \ ,
    \end{align}
while any combination 
    \begin{align}\label{KH-comb}
        k_{H}=k_{(i)}+\alpha \, k_{nonH} \ ,
    \end{align}
is non-zero at the vacuum, corresponding to a vector field being Higgsed. We can take $\alpha=1$ and expand our symplectic $(k^{u\Lambda}\,,k^u_\Lambda)$ as\footnote{Thus with the symplectic embedding above $k_H=k_1$, $k_{nonH}=k_0+k_1$.}
    \begin{align}
        k_\Lambda = \{ k_{nonH}-k_H,k_H,k_H,k_H\}\ ,
        \qquad 
        k^{u\Lambda} = \vec{0} \ ,
    \end{align}
which identifies a tensor $m_\Lambda^H$
    \begin{align}\label{k-vac-dec}
        k_\Lambda &= k_H \{-1,1,1,1\}+k_{nonH}\{1,0,0,0\}
        \nonumber \\ 
        &= k_H m^H_\Lambda + k_{nonH\,\Lambda} 
    \end{align}
where only the first term matters at the vacuum since the second, which also depends on the linear combination chosen in \eqref{KH-comb}, vanishes. 
This means in a general SUSY $AdS_4$ vacuum of N=2 supergravity in which $N$ gauge fields  are Higgsed, the symplectic Killing vector can be decomposed as
    \begin{align}
        k^u_\Lambda = \sum_{\alpha=1}^N k^{u}_{H\,\alpha} m^\alpha_\Lambda + k^u_{nonH} \ .
    \end{align}
For the $U(1)^2$-invariant truncation of N=8 we are considering, the effective theory around the vacuum thus sees of one vector fields become massive. The combination that is Higgsed can be read from the covariant derivatives of the charged scalars (the quaternionic $q^u$ coordinates) in the Lagrangian of $\mathcal{N}=2$ gauged supergravity \cite{DAuria:1990qxt}\cite{Andrianopoli:1996cm}:
    \begin{align}
        \nabla_\mu q^u = \partial_\mu q^u-g \langle A_\mu \,,\, k^u\rangle \ ,
    \end{align}
and by expanding for the effective scalars around the vacuum $\hat q^u$,
    \begin{align}
        \nabla_\mu \hat q^u = \partial_\mu \hat q^u- e A_\mu^{Higgsed}   \ ,
    \end{align}
 where $e$ is the charge of the scalar, which for this model gives
    \begin{align}
        A^{Higgsed}_\mu &= A_\mu^0-A_\mu^1-A_\mu^2-A_\mu^3 \ .
    \end{align}
By redefining the field $\tau \rightarrow t-\frac{1}{2\sqrt3}\sigma$, the effective Lagrangian for the scalar fields becomes
    \begin{align}
        \nabla_\mu  \hat q^u \nabla^\mu  \hat q^v h_{uv} =
        \frac34 \left[
                3\partial_\mu V\partial^\mu V+\partial_\mu \sigma\partial^\mu \sigma+4\partial_\mu \theta\partial^\mu \theta+6(\partial_\mu t-\frac{2}{\sqrt3}A_\mu^H)(\partial^\mu t-x\frac{2}{\sqrt3}A^{,\mu}) 
         \right]\ .
    \end{align}

\subsection{Holographic interpretation of the vacuum conditions}
To understand the extremization procedure from the field theory side we now turn to the vacuum conditions expressed in terms of the moment maps.
As reviewed in Section \ref{sec-2}, the $\tau_{RR}$ function proposed in \cite{Amariti:2015ybz} is given by 
    \begin{align}\label{tau-sugra}
        \tau_{RR}&= \frac{4F^{S^3}_{max}}{\pi^2}\frac{|W|^4}{\mathcal{I}_4}
        \ ,
        \qquad \qquad
        |W|^2 \equiv \sum_x \mathcal{L}^x\overline{\mathcal{L}}^x\ ,
    \end{align}
so that the extremization of $\tau_{RR}$ is equivalent to the extremization of $|W|^2$. The normalization of the R-charge, like in the case without hypermultiplets, is given by the vacuum condition
    \begin{align}
        \mathcal{P}^x_\Lambda s^\Lambda = c^x \ ,
        \qquad \qquad c^x c^x =1 \ .
    \end{align}
Since we've seen in eq. \eqref{Pvac} that the vacuum moment maps are actually of the form 
$
        \mathcal{P}^x_\Lambda = c^x P_\Lambda 
$,
the normalization of the R-charge in the dual SCFT can be taken to be simply 
    \begin{align}
        P_\Lambda s^\Lambda = 1 \ .
    \end{align}
In addition to this normalization, however, we have now a condition that comes from the fact that one of the Killing vector is nonzero at the vacuum. Let us look at  the structure of the Killing vectors. We have seen that they can be decomposed as in \eqref{k-vac-dec}, and this decomposition reflects on the moment maps as
    \begin{align}\label{Higgsing-P}
        \mathcal{P}^x_\Lambda = P^x_{H} m^H_\Lambda + P^x_{nH\,\Lambda} \ , 
    \end{align}
where $m_\Lambda^H$ is defined in \eqref{k-vac-dec}, $P^x_{nH\,\Lambda}=P^x_{nH}\{1,0,0,0\}$ and $(P^x_H,P^x_{nH})$ are the moment maps for the Killing vectors $(k^u_H,k^u_{nH})$. 
If we define the superpotential in the dual field theory as
    \begin{align}
        W &= P^x_\Lambda s^\Lambda =  P^x_{H} m^H_\Lambda s^\Lambda + P^x_{nH\,\Lambda}s^\Lambda 
    \end{align}
then the covariant extremization of $W$ with respect to the scalars $z^i$ and $q^u$ yields
    \begin{align}
        \mathcal{P}^{*x}_\Lambda \nabla_i s^\Lambda&=  c^x g_\Lambda \nabla_i s^\Lambda = 0 \ ,
    \end{align}
and
    \begin{align}
        \langle \nabla_uP^{x}_{H}\rangle m^H_\Lambda s^\Lambda =0 
        \qquad \Leftrightarrow \qquad 
        m^H_\Lambda s^\Lambda =0 \ .
    \end{align}
since $\nabla_u P^x_{nH}=-k^v_{nH}\Omega^x_{uv}$, and this is zero at the vacuum. It is here that we recognize how the extremization of the superpotential with respect to the quaternionic fields $q^u$ gives rise then to a linear constraint between the R-charges, for which $P^x_H$ acts as a Lagrange multiplier \cite{Amariti:2015ybz,Tachikawa:2005tq}. 

In principle, in this model we have a triplet of moment maps $\mathcal{P}^x_\Lambda$. Recall that at the vacuum 
    \begin{align}
        \mathcal{P}^x_\Lambda &= c^x g_\Lambda 
    \end{align}
where $c^x$ is an arbitrary vector on $S^2$. Thus with a global $SU(2)$ rotation we can redefine the moment map such that at the vacuum $\mathcal{P}^x_\Lambda = g_\Lambda \delta^{3,x}$. However, in this model it is not possible to extend the SU(2) transformation to a local SU(2) one that allows to  set  $\mathcal{P}^1_\Lambda=\mathcal{P}^2_\Lambda=0$ globally on the quaternionic manifold. If one is interested in the effective theory around the vacuum, actually, we can see that only one component of the SU(2) vector $\mathcal{P}^x_\Lambda$ plays a physical role.
 
To see this, let's focus on the details of the Higgsing of the $U(1)$ at the vacuum. In K\"ahler coordinates the two U(1)'s act by rotating the phases of $\zeta_1$ and $\zeta_2$ (see \eqref{U1-rotat-kahl-scalars}). The symmetry which is preserved at the vacuum is the one that rotates the phase of $\zeta_1$, in fact $\langle \zeta_1\rangle=0$. The hypermultiplet scalar charged under this isometry does not impose any constraint on the special geometry scalars, so we can study the problem restricting to the subsector where $\zeta_1=\langle \zeta_1\rangle=0$. In the quaternionic coordinates this means
    \begin{align}\label{trunc-mABJM}
        \sigma=0\ , \qquad  V=1-\theta^2-\tau^2 \ .
    \end{align}
Moreover, we expect that any constraint should also be independent on the phase of the scalar $\zeta_2$ since this is the scalar mode that's eaten by the massive gauge field.
In this subsector, it is easy to find a local SU(2) rotation that brings the moment map to the form $\mathcal{P}^x_\Lambda|_{\{\sigma=0,V+\theta^2+\tau^2=1\}}=\delta^{x,3}\mathcal{P}^3_\Lambda{}'(\theta^2+\tau^2)$.

Let's consider then the SU(2) rotated moment maps obtained from \eqref{rotP} in the decomposition analogous to \eqref{Higgsing-P}, which is relevant for the extremization at the AdS$_4$ vacuum
    \begin{align}\label{P-rot-vac-decomp}
        \mathcal{P}^{x}_{\Lambda}{}'&= P^x_{H}{}' m^H_\Lambda + P^x {}'_{nH\,\Lambda},
    \end{align}
and let's restrict to the sector \eqref{trunc-mABJM} which is consistent with our analysis around the Higgsed vacuum. The rotated moment map is 
    \begin{align}\label{P-rot-trunc}
        \mathcal{P}^{x}_{\Lambda}{}'&= \delta^{3\,x} \ 
        2(-1+\xi^2)^{-1}\{ -\xi^2 , -2+3 \xi^2 , -2+3\xi^2 , -2+3\xi^2 \}\ ,
    \end{align}
and it is a funcion of only $\xi^2=\tau^2+\theta^2$, as expected. According to the vacuum decomposition above we have
    \begin{align}
        P^x {}'_{nH\,\Lambda} &= \delta^{3\,x}\{4,0,0,0\}  \ ,
        \qquad
        P^x_{H}{}' = \delta^{3\,x} \ 2 \frac{-2+3\xi^2}{-1+\xi^2} \ .
    \end{align}
The extremization mechanism in this SU(2) gauge has the holographic normalization
    \begin{align}
        \mathcal{P}^3_\Lambda{}' s^\Lambda = 1 \ ,
    \end{align}
which, by expanding \eqref{P-rot-trunc} near $\xi=\frac{1}{\sqrt3}$ corresponds to the condition
    \begin{align}
       (s^1+3(s^2+s^3+s^4))+3\sqrt3 (s^1-s^2-s^3-s^4)\delta\xi = 1 
    \end{align}
which give the two holographic constraints
    \begin{align}\label{mABJM-constraints-sugra}
        s^1 + 3(s^2+s^3+s^4)=1\ , \qquad s^1-s^2-s^3-s^4=0\ .
    \end{align}
It emerges from the expansion above how the scalar field near the vacuum $\delta\xi$ has the role of a Lagrange multiplier for the extremization principle. The identification between these parameters in the $\tau_{RR}$ expression and the field theory charges \eqref{mABJM-charges} is analogous as for the pure ABJM case of \cite{Amariti:2015ybz} expressed through $\delta_i$ parameters. However, now one needs to additionally impose the constraints \eqref{mABJM-constraints-sugra}. This gives $\delta_0=1$ and $\delta_1+\delta_2+\delta_3=1$, which, after substituting in \eqref{tau-RR}, allows to precisely match the expression \eqref{ftres} with the supergravity prescription.

Before we move to the next example, let us comment on the hypermultiplet sector of gauged supergravities. Since the hypermultiplet structure in 4d and 5d is the same, our setup is analogous to the 5d one, with the main difference being that, instead of special geometry, 5d formulation of N=2 is given in terms of real sections, as discussed in \cite{Tachikawa:2005tq}. Moreover, the 4d $AdS_4$ vacuum solution with one hypermultiplet is analogous to the 5d one of \cite{Ceresole:2001wi}. The flow between a vacuum with a nontrivial hypermultiplet and the fully symmetric $AdS_5$ which is presented there is the 5d analogue of the 4d holographic RG flow between ABJM and its massive deformation \cite{Corrado:2001nv}.
\section{Vacua of massive type IIA theory: an ISO(7) truncation \label{sec-ISO7-trunc}}

We have considered so far the extension of the unconstrained $\tau_{RR}$ extremization (e.g. ABJM) to the case where there is a constraint due to the presence of a non-vanishing hypermultiplet scalar at the vacuum. In the setup of the previous section the theory had, like in ABJM, only electric gauging, namely the embedding tensor we considered was of the form $\Theta^M_\lambda=(0,\Theta_{\Lambda\,\lambda})$. We move, in this section, to a more general case, where the gauging is \emph{dyonic} and both $\Theta_{\Lambda\,\lambda}\neq0$, and $\Theta^{\Lambda}_\lambda\neq0$. 

The model we turn to in this section is a truncation of ISO(7) gauged supergravity \cite{Guarino:2015qaa} to the $G_0$-invariant sector for a $G_0=U(1)^2 \subset U(1)^2\times \mathbb{R}\times U(1)_{\mathbb{U}}\subset ISO(7)$ \cite{Guarino:2017pkw}, the scalars in the hypermultiplets are charged under the $\mathbb{R}\times U(1)_{\mathbb{U}}$ isometries of the hypermultiplet sector of the scalar manifold
\begin{align}
    \mathcal{M}_{sc}&= \mathcal{M}_{v}\times \mathcal{M}_{h} =\left(\frac{SU(1,1)}{U(1)}\right)^3\times \frac{SU(2,1)}{SU(2)\times U(1)}\ ,
\end{align}
which describes the scalar sector of this model. 

The special geometry in this model is derived by a square root prepotential, differently from the case of mABJM by an overall factor of $i$, and precisely
    \begin{align}
        \mathcal{F}_{mIIA}(X^\Lambda)&= -2 \sqrt{X^0X^1X^2X^3} = - i  \mathcal{F}_{mABJM}(X^\Lambda) \ ,
    \end{align}
the special geometry parametrization is 
    \begin{align}\label{SG-mIIA}
        X^0=-z^1 z^2 z^3 \ , 
        \qquad 
        X^I = - z^I \ , \qquad  I=1,2,3 \ ,
    \end{align}
giving $e^{-\mathcal{K}}=(z^1-\bar z^1)(z^2-\bar z^2)(z^3-\bar z^3)$, and $F^\Lambda(X)=\partial_\Lambda \mathcal{F}(X)$.

The gauged isometries of this theory are $SO(1,1)\times U(1)$. This is different from the previous case where the gauged isometries were both compact. We take the compact isometry to correspond to the Killing vector
\begin{align}
    k_{U(1)}=\theta\partial_\tau-\tau\partial_\theta \ ,
\end{align}
while the noncompact one is simply 
    \begin{align}
        k_{\mathbb{R}}=\partial_\sigma \ .
    \end{align}
Their corresponding moment maps, in the SU(2) gauge used in the previous subsection are
    \begin{align}\label{P-mIIA-com-noncomp}
        \mathcal{P}^x_{U(1)}=\left( -\frac{\tau}{2\sqrt{V}}\,, \frac{\theta}{2\sqrt{V}}\,,\frac{V-\theta^2-\tau^2}{4V} \right)\ ,
        \qquad
        \mathcal{P}^x_{R}=(0\,, 0\,, -\frac{1}{2V})\ .
    \end{align}
The symplectic embedding of \cite{Guarino:2015qaa} gives\footnote{This gauging corresponds to the embedding tensor
\begin{align}
    \Theta_{\mathbb{M}}^\lambda = \left(
        \begin{array}{cc}
            -c & 0 \\ \vec 0 & \vec 0 \\ g & 0 \\ \vec 0 & \vec g
        \end{array}
    \right) \ .
\end{align}
}
    \begin{align}
        k_\mathbb{M}&=g \left\{ -\frac{c}{g} \, k_{\mathbb{R}}\ , 0\ ,0\ ,0\ , k_{\mathbb{R}}\ , k_{U(1)}\ , k_{U(1)}\ , k_{U(1)}\right\}\ ,
        \nonumber
        \\
        \label{mIIA-momMaps}
        \mathcal{P}^x_\mathbb{M}&=g \left\{ -\frac{c}{g} \, \mathcal{P}^x_{\mathbb{R}}\ , 0\ ,0\ ,0\ , \mathcal{P}^x_{\mathbb{R}}\ , \mathcal{P}^x_{U(1)}\ , \mathcal{P}^x_{U(1)}\ , \mathcal{P}^x_{U(1)}\right\}
    \end{align}

In this model, we have a one-parameter family of supersymmetric $AdS_4$ vacua for 
    \begin{align}
        \langle z^i\rangle = e^{2i\pi/3}\left(\frac{c}{g}\right)^{1/3}\ ,
        \qquad
        \langle \theta\rangle =0\ , \qquad 
        \langle \tau \rangle =0\ , \qquad
        \langle V \rangle = \frac12 \left(\frac{c}{g}\right)^{2/3} \ .
    \end{align}
parametrized by $\langle\sigma\rangle\in\mathbb{R}$, where the non-compact isometry is broken and the compact one is preserved, thus in this case
    \begin{align}
        k_{H}=k_{\mathbb{R}}\ , \qquad 
        k_{nonH}=k_{U(1)} \ ,
    \end{align}
and correspondingly
    \begin{align}
        k_\mathbb{M} = (k_\Lambda,k^{\Lambda})&=k_H\{-c,0,0,0,0,g,0,0,0\}+ k_{nonH}\{ 0,0,0,0,0,1,1,1\}\ ,
        \nonumber
        \\
        &=k_{H} m_{\mathbb{M}\,H} + k_{nonH\,\Lambda}\ .
        \nonumber
        \\
    \end{align}
The relevant quantities for the holographic match is now
    \begin{align}
        \mathcal{P}^x_\mathbb{M} \mathbb{J}^{\mathbb{MN}} \mathcal{V}_\mathbb{N} = \langle \mathcal{P}^x\,, \mathcal{V}\rangle = e^{\mathcal{K}/2} \left[ 
        X^0 \mathcal{P}^x_0 - F_0 \mathcal{P}^{x\,0}-(X^1+X^2+X^3)\mathcal{P}^{x}_1\right] \ .
    \end{align}
where the moment maps can be read from \eqref{mIIA-momMaps} and \eqref{P-mIIA-com-noncomp}, while the values of the harmonic sections are given in \eqref{SG-mIIA}.

Notice that, as before, since the symmetry that rotates $\sigma,\tau$ is unbroken, we can truncate the theory to $<\theta>=<\tau>=0$. Moreover, the linearization around the vacuum won't depend on $\sigma$, since it is eaten by the gauge field becoming massive. Thus, the moment maps restricted to this subsector are   
    \begin{align}
        \mathcal{P}^{x\,,0}&=c(0,0,\frac{1}{2V})
        \ ,
        \nonumber \\
        \mathcal{P}^x_0&= - g (0,0,\frac{1}{2V})
        \ ,
        \nonumber \\
        \mathcal{P}^x_I & = g(0,0,\frac14)  \ .
    \end{align}
Notice these $\mathcal{P}^x$ are non-vanishing only in the 3-direction, so no additional $SU(2)$ rotation is required. Moreover, in order to make contact with the holographic R-charges, we can truncate the scalar fields of the vector multiplets by fixing their phase to the vacuum value
    \begin{align}\label{trunc-SG-scal-mIIA}
        z^I= e^{2/3i\pi}\rho^i\ , \qquad \rho^i\in \mathbb{R} \ .
    \end{align}
Analogously of the analysis of Section \ref{sec-U1squared-trunc}, the non-vanishing of one Killing spinor at the vacuum yields the constraint
    \begin{align}
        \langle m_{\mathbb{M}H}\,, \mathcal{V} \rangle = 0 \ .
    \end{align}
In the holographic analysis, we have to impose this constraint when we define the functional dependence between the scalars $z^1,z^2,z^3$ and the parameters $s^I$, and not just at the vacuum of the scalars $z^i$. This eliminates one of the $s^I$ and we are left with the set of the physical dual R-charges. In fact, in the holographic description, the R-charges are defined only for the U(1) fields which are not broken, which in the case are only three out of the original four vectors in the theory. Explicitly, the constraint yields
    \begin{align}
        -c F_0 + g X^0 = 0 \ , 
    \end{align}
where $F_0=\partial_{X^0}\mathcal{F}(X)$, so, from \eqref{SG-mIIA}, in terms of the scalar fields with the truncation \eqref{trunc-SG-scal-mIIA} we have the constraint
    \begin{align}\label{constr-mIIA}
        \rho^1 \rho^2 \rho^3 = \frac{c}{g} \ .
    \end{align}
Substituting this in the prepotential $\mathcal{W}$ we obtain
    \begin{align}
        \mathcal{W}= \langle \mathcal{G}_* \ , \mathcal{V}(\rho^i)\rangle|_{\rho^i\,constr.} =
        e^{-i\pi/3} \frac{g^{3/2}}{3^{3/4}c^{1/2}} (\rho^1+\rho^2+\rho^3) 
    \end{align}
where $\mathcal{G}_* = \langle\mathcal{P}^3\rangle|_{vac.}$ 
are constant coefficients given by the value of the moment maps at the vacuum, and $\rho^i$ are constraint by \eqref{constr-mIIA}. 
By eliminating $s^0$ among the set
    \begin{align}
        s^\Lambda =  e^{\mathcal{K}/2}\frac{X^\Lambda(\rho)}{\mathcal{W}} \ ,
        \qquad 
        I=0,1,2,3 \ ,
    \end{align}
thanks to the constraint, we obtain 
    \begin{align}
        |\mathcal{W}|^4&=\left( \frac{g}{c}	\right) \frac{1}{(s^1 s^2 s^3)^{4/3}} \ ,
    \end{align}
which, substituting in \eqref{tau-RR}, matches the expression \eqref{fluder}.

We presented here the analysis for the $AdS_4$ vacuum of N=2 truncation of ISO(7) gauged supergravity, but this setup can be applied in principle to more general $AdS_4$ vacua of massive IIa theory, see e.g. \cite{Passias:2018zlm}.

\section{Conclusions and Outlook \label{sec-concl}}

In this paper we have studied the supergravity dual mechanism of the minimization of the function $\tau_{RR}$ 
corresponding, in the holographic dictionary,  to the inverse square Yang Mills coupling in supergravity.
As discussed in \cite{Amariti:2015ybz} this results holds at the extremal point, while if we want to compute this 
coefficient in terms of the mixing of the graviphoton with the other vector multiplets many results
are possible.
The correct \emph{off-shell} behavior corresponding to (\ref{tau-RR}) was proposed in \cite{Amariti:2015ybz} by  studying  the case of the ABJM theory.
Here we have confirmed this expectation by studying two other truncations.
These examples, differently from the case of ABJM, have hypermultiplets and massive vector in the spectrum.
This forced us to modify the prescriptions as proposed in \cite{Amariti:2015ybz} because 
the hypermultiplets in the gravity theories play the role of the superpotential constraints on the $R$-charge of the 3D theory, 
thus playing the role of Lagrange multipliers.
Using this prescription we have reproduced the field theory results from the gauged supergravity computation in both the mABJM case and in the dual of the massive IIA truncation.

An interesting future direction is the study of the relation between the function $\tau_{RR}$
obtained from supergravity and the dual field theoretical expectations 
for  M2 branes probing seven dimensional manifolds.
An useful class of examples are the  truncations studied  in 
\cite{Cassani:2012pj}, that  correspond to seven dimensional Sasaki-Einstein manifolds 
containing non trivial 5-cycles (counted by the second Betti number of the variety). The KK reduction, performed along the Reeb vector, i.e. the isometries 
of the seven dimensional manifolds, gives origin to $AdS_4$ gauged supergravity 
with Betti vector multiplets coming from the non-trivial 5-cycles.
Furthermore there are also  massive vectors and hypermultiplets these truncations.  
These truncations in this sense are similar to the ones discussed  in this paper.
The difference is the presence of the Betti vectors, that in the putative holographic dual
field theories are associated to a set of symmetries, referred to as  baryonic.
These baryonic symmetries do not contribute to the free energy at large $N$ as discussed in
\cite{Jafferis:2011zi}, because they are related to some accidental flat directions.
Then we are left with a puzzle: on one side the gravitational off-shell $\tau_{RR}$ function depends 
to the mixing of the graviphoton with a set of 
Betti vectors, while the mixing with the vector multiplets, related to the isometries of the seven dimensional manifold, have been fixed by the  truncation; on the other side the field theoretical $\tau_{RR}$ function, 
when computable, depends on
the mixing with the flavor symmetries, dual to the isometries of the seven dimensional manifold but not on the baryonic ones. Furthermore, in many cases the large $N$ free energy of the putative dual quivers cannot 
even be computed using the rules of \cite{Jafferis:2011zi}, 
and then it is not  even clear if these models correctly reproduce the gravitational dual picture.
In some cases, where the Sasaki-Einstein manifold is toric one may try to use a different approach, that does not require the knowledge of the field theory data but just the result of the volume extracted from geometric considerations.
One can in principle compute the volumes from the toric data and  then  
associate the  charges read from geometry to the ones of  a putative field theory.
These charges are actually constrained \cite{Hosseini:2019ddy} and then one can use such constraints
 to obtain various equivalent expressions for the  volume.
Among these expressions then one should  look for the one expected from the supergravity calculation. We leave such an analysis for future investigations.

\section*{Acknowledgements}

We would like to thank Adolfo Guarino for interesting discussions. AG would like to acknowledge funding from Marie  Sk\l odowska–Curie Individual Fellowship of the European Commission Horizon 2020 Program fellowship "GaugedBH" for part of this project, and both authors would like to thank CERN Theory division for hosting them during the beginning of this project.
The work of AA has been supported in part by the Italian Ministero dell'Istruzione, 
Universit\`a e Ricerca (MIUR), in part by Istituto Nazionale di Fisica Nucleare (INFN)
through the “Gauge Theories, Strings, Supergravit\`a (GSS) research project and in
part by MIUR-PRIN contract 2017CC72MK-003.

\appendix

\section{Relating the $n_v=3$,$n_H=1$, $N=2$ model to the $U(1)^2$-invariant sector of $N=8$ supergravity \label{appendix-sugra-truncation}} %fold
\label{app-truncation}

In $N=8$ maximally gauged supergravity, the 70 scalars parametrize the fourfold anti-symmetrized vector representation of $SU(8)$, that can be decomposed in a self dual and anti-self dual part ($i,j,..=1,..,8$)
    \begin{align}
        \Sigma_{ijkl}dx^i\wedge dx^j\wedge dx^k\wedge dx^l \ ,
        \qquad 
        \Sigma_{ijkl}, \qquad 
        (\Sigma_{ijkl})^*=\epsilon_{ijklmnpq}\Sigma^{mnpq}\ .
    \end{align}
The $U(1)^2$-invariant sector can be determined by investigating the forms which are left invariant by the action of the two Cartan's generators of $SU(3)\subset SO(6)\subset SO(8)$ \cite{Warner:1983vz}. By constructing complex coordinates as
    \begin{align}\label{zN=8}
        z^1 &= x^1+ix^2\ ,\qquad z^2 =x^3+i x^4 \ ,
        \nonumber\\
        z^3&= x^5+i x^6\ , \qquad z^4=x^7+i x^8 \ ,
    \end{align}
$SU(3)$ acts as a rotation on the $z^i$'s that leaves $z^4$ invariant. Consider the forms
    \begin{align}
        J_i&=\frac i2 dz^i\wedge d\bar z^{\bar\imath} \ , \qquad i=1,2,3,4 \ ,
    \end{align}
and construct the following two- and four-froms
    \begin{align}
        &\tilde J_1^\pm=J_1-J_2-J_3\pm J_4\ ,
        \qquad
        \tilde J_2^\pm =-J_1+J_2-J_3\pm J_4\ ,
        \qquad
        \tilde J_3^\pm =-J_1-J_2+J_3\pm J_4\ ,
        \nonumber
        \\
        &\hspace{2cm} 
        F^\pm_1=\tilde J^\pm_2\wedge\tilde J^\pm_3\ ,
        \qquad 
        F^\pm_2=\tilde J^\pm_1\wedge \tilde J^\pm_3\ ,
        \qquad
        F^\pm_3=\tilde J^\pm_1\wedge \tilde J^\pm_2 \ .
        \nonumber \\
    \end{align}
Then the $U(1)^2$-invariant sector is parametrized by
    \begin{align}
        \Sigma (y_1,y_2,y_3,\omega_2,\omega_3) &=\sum_{i=1}^3 \frac14\left[ y_i (F_i^+ +F_i^-)+\bar y_i (F^+_i-F_i^-) \right]+
        \nonumber
        \\
        &\qquad +\frac14\left(\omega_2 (G_2^++i G_3^+) +\omega_3 (G_2^-+i G_3^-) + c.c.\right) \ ,
    \end{align}
where
    \begin{align}
     G_2^++iG_3^+ = dz_1\wedge dz^2\wedge dz^3\wedge dz^4\ ,
     \qquad    
     G_2^-+iG_3^- = dz_1\wedge dz^2\wedge dz^3\wedge d\bar z^4\ ,
    \end{align}
generalizing the $SU(3)$-invariant parametrization of \cite{Bobev:2010ib}. The complex $z^i$'s fields of N=2 model in Section \ref{sec-U1squared-trunc} are related to $\Sigma(y_i,\omega_2,\omega_3)$ by
    \begin{align}
        z^i=\frac{1+y_i\frac{\tanh |y_i|}{|y_i|}}{1-y_i\frac{\tanh |y_i|}{|y_i|}} \ ,
        \qquad 
        i=1,2,3 \ ,
    \end{align}
while the hypermultiplets $\zeta_1,\zeta_2$ are 
    \begin{align}
        \zeta_i=\frac{\omega_{i+1}\tanh{\sqrt{|\omega_2|^2+|\omega_3|^2}}}{\sqrt{|\omega_2|^2+|\omega_3|^2}}\ ,
        \qquad 
        i=1,2 \ .
    \end{align}
In this coordinates the metric is K\"ahler 
\begin{align}\label{Quatern-metric}
       ds^2=  g_{i\bar\jmath}d\zeta_i d\bar \zeta_{\bar\jmath} = \frac{d\zeta_1 d\bar\zeta_1+d\zeta_2d\bar \zeta_2}{1-|\zeta_1|^2-|\zeta_2|^2}+\frac{(\zeta_1 d\bar\zeta_1+\zeta_2d\bar\zeta_2)(\bar\zeta_1 d\zeta_1+\bar\zeta_2d\zeta_2)}{(1-|\zeta_1|^2-|\zeta_2|^2)^2} \ .
     \end{align}
It is useful to keep in mind the change of coordinates \cite{Ketov:2001gq}
    \begin{align}
    \zeta_1&=\frac{1-S}{1+S} \ ,
    \qquad
    \zeta_2=\frac{2C}{1+S}\ ,
    \end{align}
which, together with the K\"ahler transformation
    \begin{align}
        f(z)&=\log\left[\frac12(1+\zeta_1)\right]\ ,
    \end{align}
brings the Bergmann metric \eqref{Quatern-metric} to the form
    \begin{align}
        \label{metric-SC-coord-Univ-Hyper}
        ds^2_{Kahl}&=e^{2K} \left(
        dSd\bar S+2(S+\bar S)dC d\bar C-2(CdSd\bar C+ c.c.)    
        \right) \ ,
    \end{align}
which is a K\"ahler metric with K\"ahler potential 
$K=-\log(S+\bar S-2C\bar C)\equiv-\log(2e^{-2\phi})$.
The N=2 duality group generators are
    \begin{align}
        T^i(y_i,\bar y_{\bar\imath})& = \left(\begin{array}{cc}
            0 & y_i \\ \bar y_{\bar \imath} & 0
        \end{array}\right) \ ,
        \qquad 
        i=1,2,3 \ ,
        \qquad
        S = \left(\begin{array}{ccc}
            0 & 0 & \omega_2 \\
            0 & 0 & \omega_3 \\
            0 & \bar \omega_2 & \bar \omega_3  
        \end{array}
        \right)\ ,
    \end{align}
which parametrize $SU(1,1)^3\times SU(2,1)$. 

The $U(1)^2$ truncation of N=8 has a residual gauge symmetry which acts on the coordinates \eqref{zN=8} as $z^i\to e^{i\varphi}z^i$, $i=1,2,3$ and $z^4\to e^{i\psi}z^4$, translating into transformations of $\omega_2$ and $\omega_3$ as
    \begin{align}
        \omega_2 \to e^{i(3\varphi+\psi)}\omega_2\ ,
        \qquad 
        \omega_3 \to e^{i(3\varphi-\psi)}\omega_3
    \end{align}
as a consequence, the residual $U(1)\times U(1)$ gauge symmetry acts on the quaternionic scalars as \cite{Bobev:2010ib}
    \begin{align}\label{U1-rotat-kahl-scalars}
        \zeta_1 \to e^{i(3\varphi+\psi)}\zeta_1\ ,
        \qquad 
        \zeta_2 \to e^{i(3\varphi-\psi)}\zeta_2 \ .
    \end{align}
generated by the Killing vectors 
    \begin{align}\label{Kill-Kahl-Coord}
        k_{(i)}&=i(\zeta_1\partial_{\zeta_1}-\zeta_2\partial_{\zeta_2})+c.c. \ ,
        \qquad
        k_{(ii)}= i (\zeta_1\partial_{\zeta_1}+\zeta_2\partial_{\zeta_2})+c.c. \ .
    \end{align}
In this work we focus on the quaternionic geometry $SU(2,1)/U(2)$, which can be found in several parametrizations in the literature\footnote{In addition to \cite{Ketov:2001gq} and \cite{Ceresole:2001wi}, see for example \cite{Guarino:2017jly,Guarino:2015tja} and references therein for interesting comparisons useful to the two models discussed in this paper.}. The  quaternionic parametrization of \cite{Ceresole:2001wi} is relevant here. In particular, this is related to the real parametrization in \eqref{metric-SC-coord-Univ-Hyper} by
    \begin{align}\label{SC-to-V-coord-univ-hyper}
        S=V+(\theta^2+\tau^2)+i\sigma \ ,
        \qquad 
        C=\theta-i\tau \ .
    \end{align}
The change of coordinates from the K\"ahler ones is instead
    \begin{align}
    \theta-i\tau =  \frac{\zeta_2}{1+\zeta_1 }\ ,
        \qquad
        V=\frac{1-|\zeta_1|^2-|\zeta_2|^2}{(1+\zeta_1)(1+\bar\zeta_1)} \ ,
        \qquad 
        \sigma = \frac{i(\zeta_1-\bar\zeta_1)}{(1+\zeta_1)(1+\bar\zeta_1)} \ ,
    \end{align}
and its inverse
    \begin{align}
        \zeta_1 = \frac{1-V-(\theta^2+\tau^2)-i\sigma}{1+V+(\theta^2+\tau^2)+i\sigma} \ ,
        \qquad 
        \zeta_2 = 2\frac{\theta-i\tau}{1+V+(\theta^2+\tau^2)+i\sigma} \ .
    \end{align}
The metric in the $q^u=(V,\sigma, \theta, \tau)$ coordinates is
    \begin{align}
        ds^2&=\frac{dV^2}{2V^2}+\frac{1}{2V^2}(d\sigma+2\theta d\tau-2\tau d\theta)^2+\frac2V(d\tau^2+d\theta^2)
    \end{align}
%s
Among the two $U(1)$'s that rotate $\zeta_1$ and $\zeta_2$, one is the transformation that rotates $\theta-i\tau$ while the other is a complicated combinations of the isometries of the metric. Their embedding in $N=2$ symplectic geometry is given by 
\begin{align}
    k^u=(k^{\Lambda\,u},k_\Lambda^u) \ ,  \qquad k_\Lambda^u=(k_0^u,k_1^u,k_1^u,k_1^u) \ , \qquad k^{\Lambda u}=0 \ ,
\end{align}
with
    \begin{align}
        k_0=-k_{(i)}\ ,\qquad 
        k_1=2k_{(i)}+k_{(ii)} \ .
    \end{align}
By a change of coordinates one can check that the Killing vectors in K\"ahler coordinates \eqref{Kill-Kahl-Coord} become\footnote{It is useful to work directly with the Killing vectors and moment maps of Sec. 3.2 of \cite{Ceresole:2001wi}}
    \begin{align}\label{Kill-V-Coord}
        k^u_0 =    \left( \begin{array}{c}
            2 V \sigma \\
            1+\sigma^2-(V+\theta^2+\tau^2)^2 \\
            \sigma \theta - \tau (1+V+\theta^2+\tau^2) \\
            \sigma \tau + \theta (1+V+\theta^2+\tau^2) 
        \end{array} 
        \right) \ ,
        \qquad 
        k^u_1 = \left(
            \begin{array}{c}
               2 V \sigma \\
            1+\sigma^2-(V+\theta^2+\tau^2)^2 \\
            \sigma \theta - \tau (V+\theta^2+\tau^2-3) \\
            \sigma \tau + \theta (1+V+\theta^2+\tau^2-3) 
            \end{array}
        \right)  \ .
    \end{align}
Their corresponding moment maps are 
    \begin{align}
        \mathcal{P}^x=(\mathcal{P}^{x,\Lambda},\mathcal{P}^x_\Lambda)\ , 
        \qquad \mathcal{P}^x_\Lambda = (\mathcal{P}^x_0,\mathcal{P}^x_1,\mathcal{P}^x_1,\mathcal{P}^x_1) \ ,
        \qquad \mathcal{P}^{x\,\Lambda}=0 \ ,
    \end{align}
with
    \begin{align}\label{Pcal-V-coord}
        \mathcal{P}^x_0= -\mathcal{P}^x_{(i)} &= \left(
            \begin{array}{c}
                -\frac{2}{\sqrt V}\left(\sigma \tau+\theta(1-V+\theta^2+\tau^2)\right)\\
                -\frac{2}{\sqrt V}\left(-\sigma \theta+\tau(1-V+\theta^2+\tau^2)\right)
                \\
                -\frac{\left(
                1+V+\theta^2+\tau^2\right)^2+\sigma^2-2V(2+\theta^2+\tau^2)}{2V}
                +3 (\theta^2+\tau^2)
            \end{array}
        \right) \ ,
            \\[5mm]
        \mathcal{P}^x_1= 2 \mathcal{P}^x_{(i)} +\mathcal{P}^x_{(ii)}  &= \left(
            \begin{array}{c}
                -\frac{2}{\sqrt V}\left(\sigma \tau+\theta(-3-V+\theta^2+\tau^2)\right)\\
                -\frac{2}{\sqrt V}\left(-\sigma \theta+\tau(-3  -V+\theta^2+\tau^2)\right)
                \\
                -\frac{1}{2V}\left[(1+V+\theta^2+\tau^2)^2+\sigma^2-4(2(\theta^2+\tau^2)(1+V)-V)\right]
                \\
                \\
            \end{array}
        \right) \ ,
    \end{align}
where $\mathcal{P}^x_{(i)}\,,\,\mathcal{P}^x_{(ii)}$ refer to the Killing vectors \eqref{Kill-Kahl-Coord}.
\subsection{Local SU(2) rotation}

We consider the following SU(2) matrix, for $\tau>0$,
    \begin{align}
      U(\theta,\tau) &= \exp\left[-\frac{i}{2}\sigma^2 f(\theta,\tau) \right]\cdot  \exp\left[-\frac{i}{2}\sigma^3 g(\theta,\tau) \right]\ ,
      \nonumber \\
      & f(\theta,\tau)=\sqrt{\frac{V}{\theta^2+\tau^2}}\ ,\qquad
      g(\theta,\tau)=\frac{\theta+\sqrt{\theta^2+\tau^2}}{\tau}\ .
    \end{align}
The action of this local SU(2) rotation \eqref{P-SU2-rot} takes the moment maps above into 
    \begin{align}\label{rotP}
       2P_{H}'-3 P_{nH}'= \mathcal{P}_{(i)}'-\mathcal{P}_{(ii)}'&= \left(
            \begin{array}[]{c}
                0 \\ 0 \\- 4V^{-1}(V+\theta^2+\tau^2)
            \end{array}
        \right) \ ,
        \nonumber \\
        P_H'-\frac12 P_{nH}'=\frac32\mathcal{P}_{(i)}'+\frac12\mathcal{P}_{(ii)}'&= \left(
            \begin{array}[]{c}
           \frac{2(\theta^2+\tau^2)
            \left(\theta+\sqrt{\theta^2+\tau^2}\right) \left( -1-\sigma^2+(V+\theta^2+\tau^2)^2\right)}{\sqrt{V}(V+\theta^2+\tau^2)\left( 
                \theta^2+\tau^2+\theta\sqrt{\theta^2+\tau^2}
                \right)}   
            \\
            -\frac{4\sigma \sqrt{\theta^2+\tau^2}}{\sqrt{V}}
            \\
            -\frac{  (-V+\theta^2+\tau^2)
            \left( 1+\sigma^2 +(V+ \theta^2+\tau^2)^2 \right)}{V(V+\theta^2+\tau^2)}   
            \end{array}
        \right) \ .
    \end{align}

\section{ Definitions and useful identities of gauged N=2 supergravity\label{append-spec-geom} }

A general $N=2$ theory\footnote{For a complete discussion of $N=2$ gauged supergravity we refer to the review \cite{Andrianopoli:1996cm}.} can be coupled to $n_V$ vector multiplets $(A_\mu^I,\lambda^{iA},\lambda^{i*}_A,z^i)$, containing complex scalar fields $z^i$ $(I,i=1,..,n_V)$, and $n_H$ hypermultiplets $(\zeta^\alpha,q^u)$, containing real scalars ($\alpha=1,..,2n_H$, $u=1,...,4n_H$).
The bosonic part of the action is
\begin{align}\label{SGaction}
S&=\int d^4 x\left(- \frac R2+g_{i\bar{j}}\nabla_{\mu}z^i\nabla^{\mu}\bar z^{\bar j}+h_{uv}\nabla_\mu q^u\nabla^\mu q^v-V_g(z,\bar z,q)
\right.
\nonumber
\\
&\hspace{2cm}
\left.
+
\mathcal{I}_{\Lambda \Sigma}F^{\Lambda}_{\mu \nu}F^{\Sigma\,\mu\nu} +\frac1{2\sqrt{-g}}\mathcal{R}_{\Lambda \Sigma}\epsilon^{\mu \nu \rho \sigma}F^{\Lambda}_{\mu \nu}F^{\Sigma}_{\rho \sigma}
\right)\ .
\end{align}

\subsection{Quaternionic geometry} 
\label{subsec:quatgeo}
The $4n_H$ real $q^u$ scalars are coordinates of a quaternionic manifold $\mathcal{QM}$ of dimension dim${}_{\mathbb{Q}}=n_H$. All the models considered in this paper have a quaternionic geometry\footnote{ We refer to \cite{DAuria:1990qxt} for a detailed review on quaternionic geometry.} with positive definite metric $h_{uv}$. The choice of gauging considered in this work involves a group of isometries $G\in \mathcal{QM}$ which is always abelian. It is defined by a set of moment maps $\mathcal{P}^x(q)$ related to the Killing vectors as \cite{Louis:2012ux}
\begin{eqnarray}
    \label{eq:kill2Dmommap}
k^u_\lambda \Omega^x_{uv}&=&-\nabla_v\mathcal P^x_\lambda\ ,
\end{eqnarray}
where $\Omega^x$ is the curvature of the $SU(2)$ connection $\omega$ on the quaternionic manifold
    \begin{align}
        d\omega^x+\epsilon^{xyz}\omega^y\wedge\omega^z&=\Omega^x 
        \ ,
    \end{align}
and is covariantly constant with respect to this connection $\omega$. The moment maps and the Killing vectors satisfy an equivariant condition \cite{DAuria:1990qxt}
    \begin{align}
        \label{eq:equivariant}
        \Omega_{uv}^x k^u_\lambda k^u_\sigma = \frac12 \epsilon^{xyz}\mathcal{P}^y_\lambda \mathcal{P}^z_\sigma -\frac12 f^{\delta}_{\lambda\sigma} \mathcal{P}^x_\delta \ .
    \end{align}
Notice that, by using this identity for the curvature
    \begin{align}
        \Omega_{us}^x\Omega^y_{tw}h^{st}&=-\lambda^2\delta^{xy}h_{uw}+\lambda \epsilon^{xyz}\Omega^z_{uw}\ .
    \end{align}
It follows that
    \begin{align}
        \sum_x\Omega_{us}^x\Omega^y_{tw}h^{st}&=-3\lambda^2 h_{uw} \ ,
    \end{align}
and one can solve the expression for the Killing vectors
    \begin{align}
        \label{eq:kill2mommap}
        k^u_\lambda=\frac{1}{3\lambda^2}\sum_x h^{uw}\Omega^x_{tw}\nabla_s\mathcal{P}^x_\lambda h^{st}\ .
    \end{align}
By using \eqref{eq:kill2mommap} on the $k^v_\sigma$ in the following expression, and then the definition \eqref{eq:kill2Dmommap}, we obtain
    \begin{align}
        h_{uv}k^u_\lambda k^v_\sigma &= h_{uv}
        k^u_\lambda\left(
            \sum_x h^{vw} \Omega^x_{tw}\nabla_s \mathcal{P}^x_\sigma h^{st} 
        \right)=
        \sum_x (k^u_\lambda\Omega^x_{tu})\nabla_s\mathcal{P}^x_\sigma h^{st}=
        \nonumber
        \\
        &= -\frac1{3\lambda^2}\sum_x\nabla_u \mathcal{P}^x_\lambda h^{uv}\nabla_v \mathcal{P}^x_\sigma \ .
    \end{align}
This means that we can translate the vacuum condition 
    \begin{align}
        h_{uv}\langle k^u,\mathcal{V}\rangle\langle k^v,\overline{\mathcal{V}}\rangle &=0 \qquad \Leftrightarrow \qquad
        \sum_x h^{uv}\langle \nabla_u\mathcal{P}^x,\mathcal{V}\rangle \langle \nabla_v\mathcal{P}^x,\overline{\mathcal{V}}\rangle =0
    \end{align}
which in turn is satisfied (for a metric $h_{uv}$ with definite signature) iff 
    \begin{align}
        \langle \nabla_u\mathcal{P}^x,\mathcal{V}\rangle = 0\ .
    \end{align}
\subsubsection{Moment maps}

Once the curvature and connection are given, the moment map relative to a Killing vector $k=k^u\partial_u$ can be constructed as
    \begin{align}
        \mathcal{P}^x &= k^u\omega^x_u + W^x
    \end{align}
where $W^x$ is the compensator field for the Killing vector $k$, appearing in the Lie derivative of the curvature
    \begin{align}\label{def-compen}
        \mathcal{L}_k \Omega^x = \nabla W^x \ ,
    \end{align}
due to the local SU(2) invariance of the quaternionic K\"ahler manifold. More precisely, under an SU(2) gauge transformation parametrized by an element $U(q^u)^i_j\in SU(2)$, 
    \begin{align}
       \sum\omega^x\sigma^x= \omega \ \  &\to\ \  \omega' \ =\  U \omega U^{-1} + \frac{1}{2i}UdU^{-1}\ ,
       \nonumber
       \\
        \ \Omega  \ \ &\to \ \ \Omega' \ =\ U \ \Omega U^{-1} \ ,
    \end{align}
so from the definition \eqref{def-compen} one can derive that the compensator transforms as 
    \begin{align}
        W\ \  \to \  W' \ =  \ U W U^{-1} + k^u U\partial_\mu U^{-1} \ ,
        \end{align}
which is consistent with the transformation of the moment map\footnote{We use the notation: 
        \begin{align}\label{P-SU2-rot}
            \mathcal{P}=\sum_x\mathcal{P}^x\sigma^x 
            \ ,
            \qquad
            \mathcal{P}^x =\frac12 \textrm{Tr}[\mathcal{P}\sigma^x]\ .
        \end{align}
}
    \begin{align}
        \mathcal{P}\ \to\ \mathcal{P}'\ =\ U \mathcal{P}U^{-1} \ .
    \end{align}

Whenever $k$ refers to an abelian isometry it is possible to use this gauge invariance to set $\mathcal{P}^1=\mathcal{P}^2=0$, or equivalently to have the SU(2) matrix $\mathcal{P}$ diagonal.

\subsection{Special geometry}
The complex scalars $z^i$ parametrize a special K\"ahler manifold $\mathcal{SM}$, whose geometry is completely defined by a K\"ahler potential $K(z,\bar z)$, from which the metric of the manifold is derived as
    \begin{eqnarray}
    \label{metric}
    g_{i\bar j}(z,\bar z)&=&\partial_i\partial_{\bar j}K(z,\bar z)\ .
    \end{eqnarray}
It is convenient to parametrize the special K\"ahler scalar fields with holomorphic symplectic sections of a projective bundle, $( X^{\Lambda}(z), F_\Lambda(z))^T$ , $\Lambda=0,1,..,n_V$, satisfying (bar indicates complex conjugation)
    \begin{align}\label{FX-specKahl}
    F_\Lambda \overline X^{\Lambda}-X^{\Lambda}\overline F_\Lambda=-ie^{-K}\ .
    \end{align}
The holomorphic functions $X^\Lambda(z)$ are projective coordinates defining the special geometry complex ones, that can be chosen for example as
    \begin{align}
        z^i = \frac{X^i}{X^0} \ , \qquad i=1,...,n_V \ .
    \end{align}
In presence of a holomorphic prepotential $F(X^\Lambda)$, the functions $F_\Lambda(z^i)$ are derived from
    \begin{align}
        F_\Lambda = \partial_\Lambda F(X^\Lambda) \ ,
    \end{align}
and the K\"ahler potential is derived from \eqref{FX-specKahl} after a gauge fixing of $X^0$.
The expression \eqref{FX-specKahl} actually defines a scalar product $\langle \Omega,\overline{\Omega}\rangle$ for the symplectic vector $\Omega = (X^\Lambda,F_\Lambda)$. In general, for any $Sp(2n_v+2)$ vector $A=(A_\Lambda,A^\Lambda)$ the inner symplectic product is
\begin{eqnarray}
\langle A_1,A_2 \rangle=A_1^T\Omega A_2\ ,\qquad \Omega=\left(
\begin{array}{cc}
 0  &I_{2n_V+2} \\  -I_{2n_V+2} &0  
\end{array}
\right)\ .
\end{eqnarray}
One then obtains covariantly holomorphic symplectic sections as
    \begin{align}
        \mathcal{V}&= e^{K/2} \left( 
        \begin{array}{c}
            X^{\Lambda} \\ F_\Lambda 
        \end{array}    
        \right) \equiv
        \left(
        \begin{array}{c}
            L^\Lambda \\ M_\Lambda
        \end{array}    
        \right)
        \ , 
    \end{align}
normalized to
    \begin{align}
        \langle \mathcal{V}, \bar{\mathcal{V}} \rangle= - i\ ,
    \end{align}
whose covariant derivatives are 
\begin{eqnarray}
U_i\equiv D_i \mathcal{V}&=&\partial_i \mathcal{V}+\frac12\partial_i K\, \mathcal{V}\ ,\qquad 
 D_{\bar \imath}\mathcal{V}=\partial_{\bar \imath} \mathcal{V}-\frac12\partial_{\bar\imath} K\, \mathcal{V}=0\ ,
\nonumber\\
U_{\bar \imath}\equiv D_{\bar \imath} \bar{\mathcal{V}}&=&\partial_{\bar \imath} \bar{\mathcal{V}}+\frac12\partial_{\bar \imath} K\, \bar{\mathcal{V}}\ ,\qquad 
D_{ i}\bar{\mathcal{V}}=\partial_i \bar{\mathcal{V}}-\frac12\partial_i K\, \bar{\mathcal{V}}=0\ .
\end{eqnarray}

A useful identity of special geometry is
    \begin{align}
        \label{SG-id-1}
        \frac12 (\mathcal{M}-i\Omega) &=
        \Omega\overline{\mathcal{V}}\mathcal{V}+\Omega U_i g^{i\bar\jmath} \overline{U}_{\bar\jmath}\Omega \ ,
    \end{align}
which has been used in the paper to derive \eqref{locality}.
Moreover, by using the special geometry identity
\begin{eqnarray}
D_{\bar j}D_i\mathcal{V}=g_{i\bar j}\mathcal{V}\ ,
\end{eqnarray}
one can derive the following relations.
\begin{eqnarray}\label{HessianW}
\frac{2 \partial_{\bar j}\partial_i|W|}{|W|}\big|_{\partial_i|W|=0}&=&g_{i\bar j}\big|_{\partial_i|W|=0}\ , \nonumber\\ 
\partial_{\bar j}\partial_i |W|^{p-2}\big|_{\partial_i|W|=0}&=&g_{i\bar j}(p-2)|W|^{p-2}\big|_{\partial_i|W|=0}\ .
\end{eqnarray}

\newpage

\bibliographystyle{JHEP}
\bibliography{BibFile}

\end{document}